\documentstyle[12pt]{article}
\begin{document}

{\Large \bf REGGE CALCULUS IN THE CANONICAL FORM}\\
{\bf Vladimir Khatsymovsky}\footnote{Budker Institute of Nuclear
Physics, Novosibirsk 630090, Russia}\\
----------------------------------------------------------------------
--\\
(3+1) (continuous time) Regge calculus is reduced to Hamiltonian
form.
The constraints are classified, classical and quantum consequences
are
discussed. As basic variables connection matrices and antisymmetric
area
tensors are used supplemented with appropriate bilinear constraints.
In these variables the action can be made quasipolinomial with
$\arcsin$
as the only deviation from polinomiality. In comparison with
analogous
formalism in the continuum theory classification of constraints
changes:
some of them disappear, the part of I class constraints including
Hamiltonian one become II class (and vice versa, some new
constraints
arise and some II class constraints become I class). As a result,
the
number of the degrees of freedom coincides with the number of links
in
3-dimensional leaf of foliation. Moreover, in empty space classical
dynamics is trivial: the scale of timelike links become zero and
spacelike
links are constant.
----------------------------------------------------------------------
----\\
\newpage
\begin{center}{\bf 1.INTRODUCTION.}\end{center}

Regge calculus \cite{Regge} attracts much attention in connection
with possibility to construct quantum gravity theory free of
ultraviolet divergences. Such the possibility is usually connected
with discrete nature of the set of field variables. The latter are
link lengths of flat tetrahedrons forming piecewiseflat Regge
manifold.
To introduce canonical quantisation we need continuous time
Hamiltonian
formalism. It was studied in a number of works
\cite{Lund}-\cite{Por}.
My strategy is that of refs.\cite{Col,Brew}, in which required
formalism
is the limit of 4-dimensional Regge calculus while distances between
successive spacelike leaves tend to zero. The main problem is an
adequate choice of variables allowing one to describe in the
continuous
time limit all the degrees of freedom of an arbitrary Regge manifold
and to pass to Hamiltonian formalism in the simplest way.In refs.
\cite{Kha} tetrad-connection variables were used first considered by
Bander in ref.\cite{Ban}. In \cite{Kha} formulation of Regge
calculus
was suggested analogous to Einstein-Cartan formalism in the
continuum
general relativity (GR) and, by passing to the continuous time
limit,
Lagrangian was found, although not for quite general Regge manifold.
Using these results some trivial low-dimensional models were
considered
in refs.\cite{Kha1} illustrating possible versions of arising finite
quantum theory.

In the given paper Einstein action for arbitrary Regge manifold is
considered in the continuous time limit and reduced to the canonical
form.

\begin{center}{\bf 2.DESCRIPTION OF THE SYSTEM.}\end{center}

Our main object is Regge manifold or simplicial complex
\cite{Simpl}.
The vertices or null-di- mensional simplices $\sigma^0$ will be
denoted
by capital letters $A,~B,~C,\ldots$; $n$-simplex $\sigma^n$
(unordered)
will be denoted by also unordered sequence of it's $n+1$ vertices:
$\sigma^n=(A_{1}\ldots A_{n+1})$. $N^{(d)}_n$ is the number of
$n$-simplices in $d$-dimensional manifold (may be, infinite). In
particular, the number of $n$-simplices meeting at given $m$-simplex
$\sigma^m$ will be denoted as $N^{(d)}_n(\sigma^m)$. Now local frames
are
defined on 4-simplices $\sigma^4=(ABCDE)$. In these frames
$a,b,\ldots
=0,1,2,3$ are vector indices; metric is $\eta_{ab}={\rm
diag}(-1,1,1,1)$
and antisymmetric tensor $\epsilon^{abcd}$ corresponds to
$\epsilon^{0123}
=+1$. $A\circ B,~A*B$ are scalar and dual products of two matrices,
respectively; $^*\!B$ is dual matrix:
\begin{eqnarray}
A \circ B & \stackrel{\rm def}{=} &
\frac{1}{2}A^{ab}B_{ab}\nonumber\\
A*B & \stackrel{\rm def}{=} & A \circ (\,^{*}\!B)\\
^{*}\!B^{ab} & \stackrel{\rm def}{=} &
\frac{1}{2}\epsilon^{ab}_{~~cd}
B^{cd}\nonumber
\end{eqnarray}

\noindent In the local frames the following elements of $SO(3,1)$
are
defined: connection matrices $\Omega_{(ABCD)}$ on 3-simplices
$(ABCD)$
and curvature matrices $R_{(ABC)}$ on 2-simplices $(ABC)$. Besides,
also
antisymmetric tensors (bivectors) $v_{(ABC)}$ are defined on
2-simplices:
two vectors $l^{a}_1,l^{a}_2$ form the triangle with bivector
\begin{equation}
v^{ab}=\epsilon_{abcd}l^{c}_{1}l^{d}_{2},
\label{v=ll}
\end{equation}
whose norm $|v|\stackrel{\rm def}{=}(v\circ v)^{1/2}$ is twice the
area
of the triangle.

Einstein action for Regge manifold is written in the form
\cite{Kha}:
\begin{equation}
S=\sum_{(ABC)}|v_{(ABC)}\!|\arcsin\frac{v_{(ABC)}}{|v_{(ABC)}\!|}\circ
R_{(ABC)}.
\label{S}
\end{equation}

\noindent Here function $\arcsin$ gives angle defect on a triangle
in terms of curvature matrix $R$. The latter is product of
connection
matrices:
\begin{equation}
R_{(ABC)}=\Omega^{\varepsilon_{(ABC)D_1}}_{(ABCD_1)}\ldots
\Omega^{\varepsilon_{(ABC)D_r}}_{(ABCD_r)},
\end{equation}

\noindent where $\varepsilon_{(ABC)D}=\pm 1$ is sign function, whose
argument is pair tetrahedron $(ABCD)$ - triangle $(ABC)$. The only
requirement imposed on this function is consistency of eqs. of
motion
for $\Omega_{\sigma^3}$ which is equivalent for particular
$\Omega$'s
to closure condition for 2-surface of 3-face $\sigma^3$. This
condition
includes 2-face bivectors rotated by connection matrices required to
transform these bivectors to the same frame. In particular, in the
neighbourhood of flat space $\Omega={\bf 1}$ it takes the form
\begin{equation}
\varepsilon_{(ABC)D}v_{(ABC)}+\varepsilon_{(DAB)C}v_{(DAB)}+
\varepsilon_{(CDA)B}v_{(CDA)}+\varepsilon_{(BCD)A}v_{(BCD)}=O(\Omega-1
).
\end{equation}

\noindent Consistency of such the conditions for 3-faces of
4-simplex
$(ABCDE)$ sharing common edge $(AB)$ requires that
\begin{equation}
\varepsilon_{(ABC)D}\varepsilon_{(ABC)E}\varepsilon_{(ABD)E}
\varepsilon_{(ABD)C}\varepsilon_{(ABE)C}\varepsilon_{(ABE)D}=-1.
\label{sign-cond}
\end{equation}

Next some constraints on bivectors $v$ are required ensuring their
tetrad
structure. The difficulty is that neighbouring bivectors well may be
defined in the different frames; namely, a 4-simplex $(ABCD_0E_0)$
exists
for each $(ABC)$ where $v_{(ABC)}$ is defined (to reflect this fact
let us introduce the more detailed notation
\begin{equation}
v_{(ABC)}\equiv v_{(ABC)D_0E_0}
\end{equation}

\noindent ). Therefore it is natural to consider for each $(ABC)$ the
set
of {\bf all} $(ABCDE)$ containing this triangle and to define a
priori
arbitrary corresponding $v_{(ABC)DE}$. Now, what conditions should
be
fullfilled in order that this set of bivectors would correspond to
some
Regge manifold where these bivectors are given by (\ref{v=ll})?
First,
consider 4-simplex $(ABCDE)$ and a vertex $A$ in it. The triangles
sharing
$A$ satisfy relations on dual products of bivectors the same as those
for
bivectors in the continuum theory at a given point \cite{Kha2}:
\begin{eqnarray}
\label{v*v}
v_{(ABC)DE}*v_{(ABC)DE} & =0, & {\rm perm}(B,C,D,E)\\
\label{v*v'}
v_{(ABC)DE}*v_{(ABD)CE} & =0, & {\rm perm}(B,C,D,E)\\
\label{v*v-v*v}
\varepsilon_{(ABC)D}\varepsilon_{(ADE)C}v_{(ABC)DE}*v_{(ADE)BC}+&&\nonumber\\
\varepsilon_{(ABD)C}\varepsilon_{(ACE)D}v_{(ABD)CE}*v_{(ACE)BD} & =0,
&
{\rm perm}(B,C,D,E).
\end{eqnarray}

\noindent Second, the sum of bivectors in any tetrahedron is zero:
\begin{equation}
\varepsilon_{(ABC)D}v_{(ABC)DE}+\varepsilon_{(DAB)C}v_{(DAB)CE}+
\varepsilon_{(CDA)B}v_{(CDA)BE}+\varepsilon_{(BCD)A}v_{(BCD)AE}=0
\label{v+v}
\end{equation}

\noindent in any of two 4-simplices $(ABCDE)$ sharing the
tetrahedron
$(ABCD)$. Not all of the relations (\ref{v*v}) - (\ref{v+v}) are
independent ones since modulo (\ref{v+v}) validity of (\ref{v*v}) -
(\ref{v*v-v*v}) at any vertex $A$ means their validity at remaining
three
vertices. If (\ref{v*v}) - (\ref{v+v}) hold, tensors $v$ in the
4-simplex
are bilinears of it's edges just as analogous continuum theory
tensors
are tetrad bilinears.

Finally, third, we need conditions ensuring unambiguity of
linklengths
recovered from $v$ in the different 4-simplices. In the continuum
theory
such the problem did not exist since the tetrad was local function
of
the bivector. Now we can require continuity of scalar products of
bivectors
on 3-face $(ABCD)$ shared by 4-simplices $(ABCDE)$ and $(ABCDF)$:
\begin{eqnarray}
\label{vv-vv}
\Delta(v_{(ABC)D}\circ v_{(ABC)D})&\stackrel{\rm def}{=}&
v_{(ABC)DE}\circ v_{(ABC)DE} - v_{(ABC)DF}\circ v_{(ABC)DF}=0,
\nonumber\\{\rm perm}(A,B,C,D),&&\\
\label{vv'-vv'}
\Delta(v_{(ABC)D}\circ v_{(ABD)C})&\stackrel{\rm def}{=}&
v_{(ABC)DE}\circ v_{(ABD)CE} - v_{(ABC)DF}\circ v_{(ABD)CF} = 0,
\nonumber\\{\rm perm}(A,B,C,D).&&
\end{eqnarray}

\noindent By (\ref{v+v}) there are 6 such independent conditions for
each
3-face. This number is sufficient for continuity of it's 6 edges.
Eqs.
(\ref{vv'-vv'}), e.g., are sufficient. But the system
(\ref{vv'-vv'})
(modulo (\ref{v*v}) - (\ref{v+v})) is still highly reducible: it is
sufficient to require continuity of the edges of a triangle on only
all
but one of tetrahedrons meeting at this triangle to get continuity on
all
such tetrahedrons.

In terms of only linklengths continuous symmetries of our system are
absent since, generally speaking, any change of linklengths means
change
of geometry. Extension of the set of variables by inclusion of
connection
in our case is compensated by symmetry w.r.t. $SO(3,1)$ rotations of
local frames.

Thus, our formulation is characterised by action (\ref{S}) and by
the
system of constraints (\ref{v*v}) - (\ref{vv'-vv'}) of which we
shall
below extract irreducible ones.

\begin{center}{\bf 3.CONTINUOUS TIME.}\end{center}

Here we derive the Lagrangian. In fact, it is generalisation on
arbitrary
Regge manifold of the result of \cite{Kha} written in bivector
notations.

To pass to the continuous time let us divide the set of vertices of
Regge
manifold into 3-dimensional leaves numbered by a parameter $t$ which
we
call time and tend the step $dt$ between the leaves to zero. The
points of
the leaf will be denoted by indices $i,~k,~l,\ldots$. Let us assume
the
following consistency condition: each 4-dimensional simplex is formed
by
vertices of only two neighbouring leaves and length of one of it's
edges
is $O(dt)$. This requires for each vertex $i$ at the leaf $t$
occurence
of it's images $i^{+}$ in the leaf $t+dt$ and $i^{-}$ in the leaf
$t-dt$
such that linklengths of $(ii^{+})$ and $(i^{-}i)$ are of the order
of $dt$.
Any such 4-geometry is formed of given 3-leaf as follows. Let us
choose
any vertex $i$ and consider it's star in 3-leaf, i.e. the set of all
the
simplices of the leaf containing this vertex. Connect the image
$i^{+}$
to all the vertices of this star. Then analogous procedure can be
repeated
with the obtained "mixed" leaf (where vertex $i$ is replaced by
$i^{+}$)
and with some another vertex $k$. As a result, the leaf arises where
two
vertices are taken at $t+dt$ and others are at $t$. In analogous way
all
the rest of vertices can be shifted in time untill we get the leaf
all
points of which are taken at the time $t+dt$. It is clear that each
thus
obtained block of 4-geometry filling the space between the leaves $t$
and
$t+dt$ is specified by the consequence of the above defined time
shifts
of vertices. It is remarkable that our Lagrangian will {\bf not
depend}
on such the consequence.

To pass to the limit $dt\rightarrow 0$ let us choose sign function
$\varepsilon_{(ABC)D}$ conveniently. In 3-dimensional notations put
\begin{equation}
\varepsilon_{(i^{+}kl)i}=-1,~~~\varepsilon_{(ikl)i^{+}}=+1
\end{equation}

\noindent (this unify the form of the kinetic term)

Further, it is convenient when going to the continuous time to
assume
the continuity condition: if $(A_1A_2\ldots A_{n+1})\rightarrow
(B_1B_2\ldots B_{n+1})$ at $dt\rightarrow 0$, then $g_{(A_1A_2\ldots
A_{n+1})}\rightarrow g_{(B_1B_2\ldots B_{n+1})}$ for a quantity $g$
defined on $n$-simplices. (Convergence of one simplex to another is
understood as convergence of corresponding vertices $A_j\rightarrow
B_j$
and of vectors of links $(A_jA_k)$ and $(B_jB_k)$).

In particular, let us choose for sign function
\begin{equation}
\varepsilon_{(i^{+}kl)m}=\varepsilon_{(ikl)m},~~~
\varepsilon_{(ikl)m^{+}}=\varepsilon_{(ikl)m}.
\end{equation}

\noindent Also denote
\begin{equation}
\varepsilon_{ikl}\stackrel{\rm def}{=}\varepsilon_{(i^{+}ik)l}.
\end{equation}

\noindent Then consistency condition for sign function
(\ref{sign-cond})
is equivalent to the following one:
\begin{equation}
\varepsilon_{ikl}\varepsilon_{ikm}\varepsilon_{(ikl)m}
\varepsilon_{(ikm)l}=-1.
\end{equation}

Connection on spacelike tetrahedron at $dt\rightarrow 0$ should
describe
parallel vector transport at infinitesimal distance in time
direction
and thereby it takes the form
\begin{equation}
\Omega_{(iklm)}={\bf 1}+f_{(iklm)}dt.
\end{equation}

\noindent The same can be written for diagonal tetrahedrons with
some
vertices shifted to the next time leaf. For continuity reasons
corresponding antisymmetric matrices $f$ do not change at such shift
(as those describing vector transport at infinitesimally close
points
and at infinitesimally close directions). But this is even
inessential
since the resulting Lagrangian turns out to contain (\cite{Kha})
only
the sum $h_{(iklm)}$ of $f$'s over all four types of tetrahedrons -
$(iklm)$ and it's diagonal images with different number of vertices
shifted to the next time leaf; for example
\begin{equation}
h_{(iklm)}=f_{(iklm)}+f_{(iklm^{+})}+f_{(ikl^{+}m^{+})}+
f_{(ik^{+}l^{+}m^{+})},
\end{equation}

\noindent where antisymmetric matrix $h_{(iklm)}$ is an analog of
the
continuum GR connection $\omega_0$.

The tetrahedron connection is the discrete analog of continuum
connection
for transport orthogonal to the tetrahedron. Let us denote timelike
tetrahedron connection as
\begin{equation}
\Omega_{i(kl)}\stackrel{\rm def}{=}\Omega_{(i^{+}ikl)}
\end{equation}

\noindent (and the same for tetrahedrons differing by time shifts of
$k,~l$).

For bivectors we denote
\begin{equation}
n_{ik(lm)}dt\stackrel{\rm def}{=}v_{(i^{+}ik)lm},~~~
\pi_{(ikl)m}\stackrel{\rm def}{=}v_{(ikl)i^{+}m}
\end{equation}

Substituting the limiting form of variables into Regge action we get
the
Lagrangian where analogs of the terms $\pi\dot{\omega}$, $h{\cal
D}_{\alpha}
\pi^{\alpha}$ and $n_{\alpha}R^{\alpha}$ of the continuum theory
\cite{Kha2}
can be viewed denoted below $L_{\dot{\Omega}}$, $L_h$ and $L_n$,
respectively.
Besides, some new terms appear due to the difference of limiting
curvature
matrices $R$ on spacelike and diagonal triangles from unity\footnote
{The closure of these $R$ to unity would be natural to assume for
their
contribution to $L$ be finite \cite{Brew}. However, the finiteness
can
be achieved at noninfinitesimal $R-1$ as well on condition that
contributions
of neighbouring triangles cancel each other, just as in this work.}.
Indeed, write out the finite part of curvature matrix $R_{(ikl)}$
if,
e.g., triangle $(ikl)$ is common 2-face of the timelike tetrahedrons
$(i^{-}ikl)$ and $(ik^{+}kl)$:
\begin{equation}
R_{(ikl)}=\Omega^{\varepsilon_{(ikl)i^{-}}}_{(i^{-}ikl)}
\Omega^{\varepsilon_{(ikl)k^{+}}}_{(ik^{+}kl)}+O(dt)=
\Omega^{\dag}_{i(kl)}\Omega_{k(li)}+O(dt).
\end{equation}

\noindent Normals to the tetrahedrons $(i^{-}ikl)$ and $(ik^{+}kl)$
are, generally speaking, different, just as vectors of links
$(i^{-}i)$
and $(kk^{+})$ are (the latter being analogs of shift-lapse functions
at
different points), so $\Omega_{i(kl)}$ and $\Omega_{k(li)}$ do not
necessarily coincide. These matrices, however, are not quite
independent
as follows from the equations of motion for connection; relation
between
them will also ensure finiteness of the Lagrangian. Indeed, at the
infinitesimal variation
\begin{equation}
\delta\Omega_{k(li)}=w_{k(li)}\Omega_{k(li)}dt,~~~w^{\dag}_{k(li)}=
-w_{k(li)}
\end{equation}

\noindent finite addends to the Lagrangian will arise only from
potentially infinite terms (contribution of the triangles with
noninfinitesimal area and defects). There are two such terms
containing
$\Omega_{k(li)}$ - contributions of $R_{(ikl)}$ and $R_{(ik^{+}l)}$.

Resulting variation of $L$ is linear in $w_{k(li)}$ and leads to a
constraint on $\Omega,~\pi$. Permuting $i,~k,~l$ we get the system
which is solvable \cite{Kha} and gives
\begin{equation}
\Omega_{i(kl)}=\Omega_{(ikl)}\exp(\phi_{i(kl)}\pi_{(ikl)}+
\,^{*}\!\phi_{i(kl)}\,^{*}\!\pi_{(ikl)}),
\end{equation}

\noindent where $\phi_{i(kl)},\,^{*}\!\phi_{i(kl)}$ are parameters.
Noninfinitesimal contribution of $(ikl)$ into action (and thus
infinite
one into $L$) is proportional to $\phi_{k(li)}-\phi_{i(kl)}$.
Contributions of diagonal triangles differs by cyclic permutations
of
$i,~k,~l$ so that the sum vanishes, e.g. (see Fig.1)
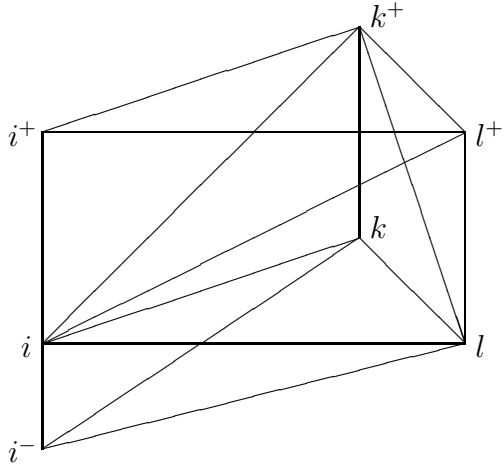
\begin{figure}
\label{prism}
\begin{picture}(200,200)(0,20)
\put (150,20){\line(0,1){120}}
\put (150,20){\line(3,2){120}}
\put (150,20){\line(4,1){160}}
\put (150,60){\line(1,1){120}}
\put (150,60){\line(2,1){160}}
\put (150,60){\line(3,1){120}}
\put (150,60){\line(1,0){160}}
\put (150,140){\line(3,1){120}}
\put (150,140){\line(1,0){160}}
\put (270,100){\line(0,1){80}}
\put (270,100){\line(1,-1){40}}
\put (270,180){\line(1,-1){40}}
\put (270,180){\line(1,-3){40}}
\put (310,60){\line(0,1){80}}
\put (310,135){$~l^{+}$}
\put (270,180){$~k^{+}$}
\put (137,135){$i^{+}$}
\put (142,55){$i$}
\put (270,100){$~k$}
\put (310,55){$~l$}
\put (137,15){$i^{-}$}
\end{picture}
\caption{Infinitesimal 3-prism}
\end{figure}
\begin{equation}
\phi_{k(li)}-\phi_{i(kl)}+\phi_{l(ik)}-\phi_{k(li)}+
\phi_{i(kl)}-\phi_{l(ik)}=0
\end{equation}

\noindent Important is that these differences are to be multiplied
by
close up to $O(dt)$ areas of images of $(ikl)$. In the next in $dt$
order to get finite terms in the Lagrangian (below denoted as
$L_{\phi}$)
one should take into account infinitesimal area differences. The
latter
depend on $n$, the lateral (timelike) 2-face bivectors.

The resulting Lagrangian reads
\begin{eqnarray}
\label{L-Regge}
L_{\rm Regge} & = & L_{\dot{\Omega}}+L_h+L_n+L_{\phi}\\
L_{\dot{\Omega}} & = & \sum_{(ikl)} \pi_{(ikl)}\circ
{\Omega}^{\dag}_{(ikl)}
\dot{\Omega}_{(ikl)}\\
L_h & = & \sum_{(iklm)}h_{(iklm)}\circ\sum_{{\rm cycle\, perm}\,
iklm}
\varepsilon_{(ikl)m}\Omega^{\delta_{(ikl)m}}_{(ikl)}\pi_{(ikl)}
\Omega^{-\delta_{(ikl)m}}_{(ikl)}\\
(\delta & \stackrel{\rm def}{=} &
\frac{1+\varepsilon}{2})\nonumber\\
L_n & = & \sum_{ik}|n_{ik}\!|\arcsin\frac{n_{ik}}{|n_{ik}\!|}\circ
R_{ik}\\
(R_{ik} & = & \Omega^{\varepsilon_{ikl_n}}_{i(kl_n\!)}\ldots
\Omega^{\varepsilon_{ikl_1}}_{i(kl_1\!)},~~~ \varepsilon_{ikl_j}\!=
\varepsilon_{(ikl_j)l_{j-1}}\!=-\varepsilon_{(ikl_j)l_{j+1}})\nonumber
\\
L_{\phi} & = & -\sum_{(ikl)}\pi_{(ikl)m}\circ\sum_{{\rm perm}\, ikl}
\varepsilon_{ikl}\phi_{i(kl)}n_{ik(lm)}
\end{eqnarray}

\noindent (entering last equation scalar products $\pi_{(ikl)m}\circ
n_{ik(lm)}$ do not depend on $m$ due to the further considered
continuity
of scalar products of bivectors). Appearing here in kinetic term
bivector
 $\pi_{(ikl)}$ is $\pi_{(ikl)m}$ at $m=m_-(ikl)$, i.e. it is
bivector
of a triangle $(ikl)$ defined in the one of two tetrahedrons
$(iklm_{\pm})$
with the face $(ikl)$ in 3-leaf whose vertices $m_+(ikl), m_-(ikl)$
(functions of $(ikl)$) are defined according to $\varepsilon_{(ikl)
m_{\pm}}=\pm 1$. We shall also write $\pi_{(ikl)\pm}$ or simply
$\pi_{\pm}$
for corresponding bivectors. Bivector $n_{ik}$ is $n_{ik(lm)}$ for
some
$(lm)$. Thus, $\pi_-,~\Omega$ are dynamical variables.

For varying in $\phi,\,^{*}\!\phi,~\Omega$ let us introduce matrices
$U=\exp(\phi\pi+\,^{*}\!\phi\,^{*}\!\pi)$, so that
\begin{equation}
\Omega_{i(kl)}=\Omega_{(ikl)}U_{i(kl)},~~~\phi=\frac{1}{|\pi|}
\arcsin\frac{\pi}{|\pi|}\circ U.
\end{equation}

\noindent It is convenient to treat $\Omega,~U$ as matrices of
general
form and take into account the conditions of orthogonality and
required
dependence of $U$ on $\pi$ with the help of Lagrange multipliers
by adding to $L_{\rm Regge}$ the following terms:
\begin{eqnarray}
\label{L-rot}
L_{\rm rot}=\sum_{(ikl)}B_{(ikl)}\circ (\Omega^{\dag}_{(ikl)}
\Omega_{(ikl)}-{\bf 1}) & + & \sum_{i(kl)}P_{i(kl)}\circ (U_{i(kl)}
\pi_{(ikl)}U^{\dag}_{i(kl)}-\pi_{(ikl)})\nonumber\\
 & + & \sum_{i(kl)}M_{i(kl)}
\circ (U^{\dag}_{i(kl)}U_{i(kl)}-{\bf 1}).
\end{eqnarray}

\noindent Lagrange multipliers are symmetric ($B,~M$) and
antisymmetric
($P$) matrices.

It remains to add to $L_{\rm Regge}$ with the help of Lagrange
multipliers constraints on bivectors (\ref{v*v}) - (\ref{vv'-vv'})
where
we shall pass to the notations $\pi, n$ and extract irreducible
constraints. Conditions on the dual products $v*v^{\natural}$ where
$v^{\natural}$ is $v$ or $v^{\prime}$ result in the constraints
$\pi *\pi^{\natural},~\pi *n$ and $n*n^{\natural}$. Since algebraic
sum
of $\pi$ in the tetrahedron in 3-leaf is zero, there are 6
independent
constraints $\pi *\pi^{\natural}$ in the tetrahedron. The number of
constraints $\pi *n$ and $n*n^{\natural}$ is 8 and 6, respectively,
in the
tetrahedron at each vertex whose shift-lapse vector form given
$n$'s.

The closure condition (\ref{v+v}) for 3-leaf tetrahedrons reads
\begin{equation}
\label{pi+pi}
\varepsilon_{(ikl)m}\pi_{(ikl)m}+\varepsilon_{(mik)l}\pi_{(mik)l}+
\varepsilon_{(lmi)k}\pi_{(lmi)k}+\varepsilon_{(klm)i}\pi_{(klm)i}=0.
\end{equation}

\noindent For the timelike tetrahedrons conditions (\ref{v+v}) allow
us
to express variations of bivectors $\pi$ due to time shift of any
vertex of
3-leaf in terms of bivectors $n$. These conditions were already used
to express variations of $\pi$ appearing when finding $L_{\phi}$.

Subtracting from the number of components of $\pi,~n$ (which is
$96N^{(3)}_3$) the number of constraints (\ref{pi+pi}) and of those
of
$v*v^{\natural}$ type gives
\begin{equation}
28N^{(3)}
\label{28N}
\end{equation}

\noindent for the number of 4-prism parameters. This is natural
since
any 4-prism is defined by 22 linklengths; in addition, there are 6
rotational degrees of freedom.

Continuity conditions for scalar products (\ref{vv-vv}) and
(\ref{vv'-vv'})
also take different form on spacelike and timelike 3-faces. Namely,
continuity on spacelike (and diagonal) faces means constraints {\bf
with
derivatives}, whose existence might change dynamical content of
theory
apart from being simply analog of continuum GR. Fortunately, as it
is
proved below, {\bf given dynamical constraints are consequences of
the
equations of motion for Lagrangian $L_{\rm Regge}$ and can be
omitted}

For example, consider continuity of the values like $v\circ
v^{\natural}$
on 3-face $(ik^{+}lm)$ (see Fig.2).
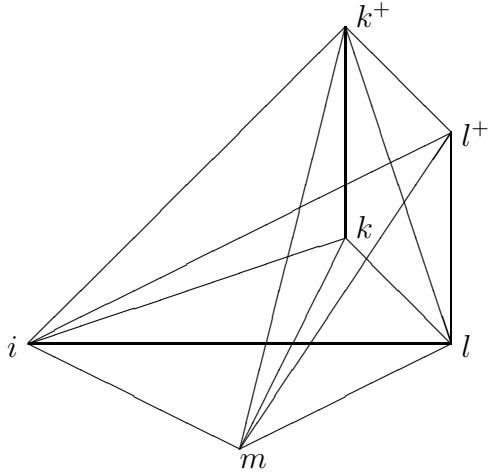
\begin{figure}
\label{face}
\begin{picture}(200,200)(0,20)
\put (150,60){\line(1,1){120}}
\put (150,60){\line(2,1){160}}
\put (150,60){\line(3,1){120}}
\put (150,60){\line(1,0){160}}
\put (150,60){\line(2,-1){80}}   
\put (270,100){\line(0,1){80}}
\put (270,100){\line(1,-1){40}}
\put (270,180){\line(1,-1){40}}
\put (270,180){\line(1,-3){40}}
\put (310,60){\line(0,1){80}}
\put (230,20){\line(1,4){40}}   
\put (230,20){\line(1,2){40}}   
\put (230,20){\line(2,3){80}}   
\put (230,20){\line(2,1){80}}   
\put (310,135){$~l^{+}$}
\put (270,180){$~k^{+}$}
\put (142,55){$i$}
\put (270,100){$~k$}
\put (310,55){$~l$}
\put (230,13){$m$}
\end{picture}
\caption{Diagonal 3-face $(ik^{+}lm)$ -
common for 4-simplices $(ikk^{+}lm)$ and $(ik^{+}ll^{+}m)$.}
\end{figure}

\noindent The difference between bivectors of close spacelike
(diagonal) 2-simplices {\bf in the same frame} on shifting the
vertex $k$
\begin{eqnarray}
D_k\pi_{(ikl)m}/dt & \stackrel{\rm def}{=} & (v_{(ik^{+}l)km}-
v_{(ikl)k^{+}m})/dt\nonumber\\ & = & \varepsilon_{kli}n_{kl(im)}-
\varepsilon_{kil}n_{ki(lm)},
\end{eqnarray}

\noindent which is an analog of {\bf covariant} derivative
(in fact, already used when finding $L_{\phi}$). The difference
of bivectors of the same 2-simplices {\bf in different frames}
(an analog of usual derivative)
\begin{equation}
\delta_k\pi_{(ikl)m}\stackrel{\rm def}{=}v_{(ik^+l^+)km}-
v_{(ikl)k^+m}.
\end{equation}

\noindent The continuity conditions connect $\delta\pi$ and $D\pi$:
\begin{eqnarray}
\label{pi-delta-pi}
\pi_{(ikl)m}\circ (\delta_i\pi_{(ikl)m}-D_i\pi_{(ikl)m})&=&0,~~~{\rm
perm}~~i,~k,~l;\\
\label{pi-delta-pi'}
\pi_{(ilm)k}\circ
(\delta_i\pi_{(ikl)m}-D_i\pi_{(ikl)m})&&\nonumber\\
+\pi_{(ikl)m}\circ (\delta_i\pi_{(ilm)k}-D_i\pi_{(ilm)k})&=&0,~~~{\rm
perm}~~i,~k,~l.
\end{eqnarray}

However, $\delta\pi_{(ikl)m}$ enter equations of motion for $L_{\rm
Regge}$
only in the form of full derivative (and for such $m$ that
$\varepsilon_{(ikl)m}=-1$)
\begin{equation}
\dot{\pi}_{(ikl)}dt=(\delta_i+\delta_k+\delta_l)\pi_{(ikl)}.
\end{equation}

\noindent Therefore constraints (\ref{pi-delta-pi}),
(\ref{pi-delta-pi'})
are equivalent to relations between $\dot{\pi}$ and $D\pi_{(ikl)}=
(D_i+D_k+D_l)\pi_{(ikl)}$:
\begin{eqnarray}
\label{pi-d-pi}
\pi_{(ikl)m}\circ (\dot{\pi}_{(ikl)m}-D\pi_{(ikl)}/dt) & = & 0\\
(D\pi_{(ikl)m}/dt=\sum_{{\rm perm}\,
ikl}\varepsilon_{ikl}n_{ik(lm)}),
\nonumber\\
\label{pi-d-pi'}
\pi_{(ilm)k}\circ (\dot{\pi}_{(ikl)m}-D\pi_{(ikl)m}/dt)\nonumber\\
+\pi_{(ikl)m}\circ (\dot{\pi}_{(ilm)k}-D\pi_{(ilm)k}/dt) & = & 0.
\end{eqnarray}

Equations (\ref{pi-d-pi}), (\ref{pi-d-pi'}) were earlier said to be
consequences of the equations of motion for Lagrangian $L_{\rm
Regge}$
supplemented with the rest of constraints on bivectors (without
derivatives). Indeed, (\ref{pi-d-pi}) at $\varepsilon_{(ikl)m}=-1$
arises
immediately from $L_{\rm Regge}$ at the following variation of
connection
type variables:
\begin{eqnarray}
\label{xi}
\Omega_{(ikl)} & \rightarrow & \Omega_{(ikl)}\exp
(\xi_{(ikl)}\pi_{(ikl)}),\nonumber\\
\phi_{i(kl)} & \rightarrow & \phi_{i(kl)}-\xi_{(ikl)},
{}~~~{\rm perm}~~i,~k,~l.
\end{eqnarray}

\noindent Namely, the constraint (\ref{pi-d-pi}) turns out to be
added to
Lagrangian multiplied by $-\xi_{(ikl)}$. Further, there is area
continuity
condition $|\pi_{(ikl)m_-}\!|^2=|\pi_{(ikl)m_+}\!|^2$ among the
scalar
product continuity constraints. Differentiating it will lead to
(\ref{pi-d-pi}) also at $\varepsilon_{(ikl)m}=+1$. Besides, in the
nondegenerate Regge manifold $N^{(3)}_2\geq N^{(3)}_1$ (this follows
from
simple combinatorial discussion with taking into account the fact
that
each edge is shared by no less then three 2-faces). Therefore all
the
links in 3-leaf and, in turn, scalar products of different bivectors
$\pi$
can be expressed through triangle areas. Corresponding relations of
the
type $\pi\circ\pi^{\prime}=f(|\pi|)$ should follow from the below
written
irreducible set of linear and bilinear constraints on bivectors
$\pi,~n$
corresponding to Regge manifold. Since we already know how to
differentiate
areas, using these relations will give derivatives
$d(\pi\circ\pi^{\prime})/dt$ in terms of $\pi,~n$. The obtained
relations
are purely kinematic ones valid for arbitrary Regge manifold and
therefore
these are no else than (\ref{pi-d-pi'}).

Thus we have shown that kinematic constraints with derivatives
(\ref{pi-d-pi}), (\ref{pi-d-pi'}) follows from equations of motion
for Lagrangian supplemented with constraints without derivatives and
should be omitted.

It remains to separate out irreducible conditions of continuity of
scalar products (\ref{vv-vv}), (\ref{vv'-vv'}) on timelike 3-faces.
In 3-leaf this corresponds to continuity on (spacelike) triangles. On
the
triangle $(ikl)$ shared by tetrahedrons $(iklm)$, $(ikln)$ we have
for the
bivectors of timelike and spacelike triangles meeting at vertex $i$:
\begin{eqnarray}
\label{Delta-pi-pi}
\Delta(\pi_{(ikl)}\circ\pi_{(ikl)}) & \stackrel{\rm def}{=} &
\pi_{(ikl)m}\circ\pi_{(ikl)m}-\pi_{(ikl)m}\circ\pi_{(ikl)m}=0,\\
\Delta(\pi_{(ikl)}\circ n_{ik}) & = & 0,\\
\Delta(\pi_{(ikl)}\circ n_{il}) & = & 0,\\
\Delta(n_{ik}\circ n_{ik}) & = & 0,\\
\Delta(n_{il}\circ n_{il}) & = & 0,\\
\label{Delta-n-n'}
\Delta(n_{ik}\circ n_{il}) & = & 0.
\end{eqnarray}

\noindent By permutations of $i,~k,~l$ we get additionally 5
analogous
equations at vertices $k,~l$ (equation (\ref{Delta-pi-pi}) remains
unchanged). Continuity of edges of tetrahedron $(ii^+kl)$ and, in
particular,
of the triangle $(ikl)$ follows from (\ref{Delta-pi-pi}) -
(\ref{Delta-n-n'}).
But continuity of the triangle $(ikl)$ follows also from the
constraints at
vertices $k,~l$ as well, that is, some constraints are superfluous.
In any
case, it is sufficient to keep constraints $\Delta(\pi\circ n)$ at
only one
vertex of each triangle which gives their full number $2N^{(3)}_2$.
The
latter is also abundant: the constraint $\Delta(\pi_{(ikl)}\circ
n_{ik})$
at all others fullfilled expresses continuity of the length of $(ik)$
in
3-leaf which should be stated on all {\bf but one} triangles $(ikl)$
meeting at this edge. Their full number thus becomes
$2N^{(3)}_2-N^{(3)}_1$.
The constraint $\Delta(n_{ik}\circ n_{il})$ can be associated with
length
continuity of edge $(ii^+)$. It suffices to impose it on
$N^{(3)}_3(i)-1$
meeting at $i$ triangles $(ikl)$. Summation over vertices gives
$4N^{(3)}_3-
N^{(3)}_0$ for the number of independent constraints of this type.
Finally,
constraints $\Delta(n_{ik}\circ n_{ik})$ are given on all but one
triangles
meeting at (ordered) 1-simplex $(ik)$; their full number is
$6N^{(3)}_2-
2N^{(3)}_1$. The number of constraints
$\Delta(\pi_{(ikl)}\circ\pi_{(ikl)})$
is $N^{(3)}_2$. As a result, the full number of constraints of the
type of
$\Delta(v\circ v^{\natural})$ is $22N^{(3)}_3-3N^{(3)}_1-N^{(3)}_0$
(with
taking into account that $N^{(3)}_2=2N^{(3)}_3$). This should be
subtracted
from (\ref{28N}) to give
\begin{equation}
6N^{(3)}_3+3N^{(3)}_1+N^{(3)}_0
\end{equation}

\noindent degrees of freedom. Of this number in each 3-leaf
$6N^{(3)}_3$
is the number of parameters of local rotations, $N^{(3)}_0$ is the
number
of timelike lengths while $N^{(3)}_1$ is the number of spacelike
ones;
since we consider block of 4-geometry between the two leaves, we take
into
account here the number of spacelike edges in two leaves,
$2N^{(3)}_1$,
plus the number of diagonal edges $N^{(3)}_1$. However, when we glue
different blocks together, we need $N^{(3)}_1$ continuity conditions
on
3-leaf between them. These are just conditions contained in
(\ref{pi-d-pi}),
(\ref{pi-d-pi'}) and shown above to follow from equations of motion.
As
a result, we have $2N^{(3)}_1+N^{(3)}_0$ independent linklengths at
arbitrary time as it should be for the Regge 4-manifold constructed
of the most general 3-leaf.

The constraints introduced, $v*v^{\natural}$ and $\sum v,~~
v\circ v^{\natural}$, can be taken into account with the help of
Lagrange multipliers by adding to Lagrangian the terms $L_{\rm
dual}$
and $L_{\rm scal}$, respectively:
\begin{eqnarray}
\label{L-dual}
L_{\rm
dual}&=&\sum_{(ik)(lm)}\!^{*}\!\mu_{(ik)(lm)}\pi_{(ikl)m}*\pi_{(ikm)l}
+\sum_{i(kl)m}\!^{*}\!\nu_{i(kl)m}n_{ik(lm)}*n_{il(km)}\nonumber\\
&+&\sum_{ik(lm)}\!^{*}\!\nu_{ik(lm)}n_{ik(lm)}*n_{ik(lm)}
+\sum_{iklm}\!^{*}\!\lambda_{iklm}\pi_{(ikl)m}*n_{ik(lm)}\nonumber\\
&+&\sum_{ik(lm)}\!^{*}\!\lambda_{ik(lm)}[\varepsilon_{ilk}
\varepsilon_{imk}\pi_{(ilm)k}*n_{ik(lm)}-\,^{*}\!\Lambda_{i(klm)}],\\
\label{L-scal}
L_{\rm scal}&=&\sum_{(iklm)}\eta_{(iklm)}\circ\sum_{{\rm cycle\,
perm}\,
iklm}\varepsilon_{(ikl)m}\pi_{(ikl)m}+\sum_{(ikl)m}\mu_{(ikl)m}
(\pi_{(ikl)m}\circ\pi_{(ikl)m}-A_{(ikl)})\nonumber\\
&+&\sum_{ik(lm)}\nu_{ik(lm)}(n_{ik(lm)}\circ n_{ik(lm)}-\sigma_{ik})
+\sum_{iklm}\chi_{_1}(ikl)\lambda_{iklm}(\pi_{(ikl)m}\circ
n_{ik(lm)}
-\Lambda_{ikl})\nonumber\\
&+&\sum_{i(kl)m}\chi_{_S}(i(kl))[\lambda_{iklm}(\pi_{(ikl)m}\circ
n_{ik(lm)}
-\Lambda_{ikl})+\lambda_{ilkm}(\pi_{(ikl)m}\circ n_{il(km)}
-\Lambda_{ilk})]\nonumber\\
&+&\sum_{i(kl)m}\chi_{_\Sigma}(i(kl))\nu_{i(kl)m}(n_{ik(lm)}\circ
n_{il(km)}-\sigma_{i(kl)}).
\end{eqnarray}

\noindent Here
$^*\!\mu,~~^*\!\nu,~~^*\!\lambda,~~\mu,~~\nu,~~\lambda,~~
\eta$ and also $^*\!\Lambda,~~A,~~\Lambda,~~\sigma$ are sets of
Lagrange
multipliers; $\chi_{_1},~~\chi_{_S},~~\chi_{_\Sigma}$ are
characteristic
functions of some sets of simplices $S_1,~~S,~~\Sigma$, arising at
constructing irreducible set of constraints above. The $S,~~\Sigma$
are sets of 2-simplices with marked vertex, on each of which 2
constraints
$\Delta(\pi\circ n)$ and(or) 1 constraint $\Delta(n\circ n^{\prime})$
are
set, respectively. $S_1$ is the set of 2-simplices with marked both
vertex
and edge on which 1 constraint $\Delta(\pi\circ n)$ is set. It is
convenient that the sets $S,~~\Sigma$ be chosen so that continuity
of
$\pi_{(ikl)}\circ n_{ik},~~\pi_{(ikl)}\circ n_{il}$ and $n_{ik}\circ
n_{il}$ on necessary number of triangles were fullfilled
simultaneously
in order that continuity of edges on these triangles would follow
immediately. I have check possibility of such choice for two simple
examples of 3-leaf: the simplest periodic Regge manifold \cite{Roc}
and
simplest closed one - 3-surface of the 4-simplex.

As a result, quasipolinomial Lagrangian of Regge calculus takes the
form of the sum of expressions (\ref{L-Regge}), (\ref{L-rot}),
(\ref{L-dual}) and (\ref{L-scal}):
\begin{equation}
\label{L-full}
L=L_{\rm Regge}+L_{\rm rot}+L_{\rm dual}+L_{\rm scal}
\end{equation}

\begin{center}{\bf 4.THE STRUCTURE OF CONSTRAINTS.}\end{center}

Proceeding to discussion of dynamics, consider full time derivative
of
some quantity $f$ in the system with Lagrangian (\ref{L-full}) which
can
be written symbolically as
\begin{equation}
L=\pi\circ {\Omega}^{\dag}\dot{\Omega}-H.
\end{equation}

\noindent Here $H$ is function of $\pi,~\Omega$ and other variables.
If
$f$ is function of $\pi,~\Omega$ then it follows with the help of
equations of motion that
\begin{equation}
\frac{df}{dt}=\{f,H\},
\end{equation}

\noindent where Poisson brackets for specific form of the kinetic
term
in $L$ prove to be
\begin{equation}
\label{{}}
\{f,H\}=\pi\circ [H_{\pi},f_{\pi}]+H_{\pi}\circ {\Omega}^{\dag}
f_{\Omega}-f_{\pi}\circ {\Omega}^{\dag}H_{\Omega}.
\end{equation}

\noindent Here indices $\pi,~\Omega$ mean corresponding derivative,
which over $\pi$ is assumed to be antisymmetrised.

The Hamiltonian $H$, as in the continuum theory, turns out to be
linear
combination of constraints, i.e. it vanishes on their surface.
Indeed,
for $L_{\rm dual},~~L_{\rm scal},~~L_{\rm rot}$ it is so by
construction.
It is also evident for $L_h$, while $L_n+L_{\phi}$ is the sum over
vertices
of the groups of terms $-H_i$ each of which is uniform function of
degree 1 of the set $n_{ik}$ for all possible $k$ at given $i$. The
$n$'s
of this set can be multiplied by some general factor without
violating
other constraints. This variation leads to Hamiltonian constraint
$H_i$:
\begin{equation}
\label{Ham}
L_n+L_{\phi}=-\sum_i H_i,~~H_i=0.
\end{equation}

\noindent Requiring the constraints be conserved in time allows us
to define Lagrange multipliers. Those at II class constraints are
defined
uniquely and therefore in the absence of matter are zero. Therefore
classical dynamics is governed in this case by I class constraints.

Proceeding to classification of constraints let us first establish
continuous symmetries. The latter correspond to occurence of I class
constraints. Originally in the full discrete theory we have symmetry
w.r.t. $SO(3,1)$ rotations in the local frames in 4-simplices. In
the
continuous time limit we have rotations in 4-prisms or, equivalently
to
say, in their tetrahedron bases; also we have some transformations
of
 $\phi,\,^{*}\!\phi$. Tetrahedron rotations $U_{(iklm)}\in SO(3,1)$
result in
\begin{eqnarray}
v & \rightarrow & U_{(iklm)}vU^{\dag}_{(iklm)},\\
\Omega_{(ikl)} & \rightarrow &
(U_{(iklm)}\Omega^{\varepsilon_{(ikl)m}}
_{(ikl)}U^{\dag}_{(ikln)})^{\varepsilon_{(ikl)m}},\\
h_{(iklm)} & \rightarrow & U_{(iklm)}h_{(iklm)}U^{\dag}_{(iklm)}-
\dot{U}_{(iklm)}U^{\dag}_{(iklm)},
\end{eqnarray}

\noindent where $v$ is bivector $\pi$ or $n$ in the tetrahedron
$(iklm)$;
$(ikln)$ is another tetrahedron in 3-leaf with the same 2-face
$(ikl)$.
It is easy to check that on functions of $\pi,\Omega$ infinitesimal
rotations $U=1+u$ are generated by Gaussian constraint
$C(u)=-L_h|_{h
\rightarrow u}$ by means of Poisson brackets $\{C(u),\cdot\}$
(see (\ref{{}})).

The invariance at shifts $\phi,~^{*}\!\phi$ is due to ambiguity when
dividing $\Omega_{i(kl)}$ into symmetric in $i,~k,~l$ part and
rotation
$\exp (\phi\pi+\,^{*}\!\phi\,^{*}\!\pi)$ not changing $\pi_{(ikl)}$.
In
particular, symmetry transformations at shift $\,^{*}\!\phi$ take the
form
\begin{eqnarray}
^{*}\!\phi_{i(kl)} & \rightarrow &
\,^{*}\!\phi_{i(kl)}-\zeta_{(ikl)},\\
\Omega_{(ikl)} & \rightarrow & \Omega_{(ikl)}\exp
(\zeta_{(ikl)}\,^{*}
\!\pi_{(ikl)}),\\
\,^{*}\!\mu_{(ik)(lm)} & \rightarrow &
\,^{*}\!\mu_{(ik)(lm)}+\frac{1}{2}
\dot{\zeta}_{(ikl)}, \ldots {\rm cycle~perm}~~i,~k,~l \ldots
\end{eqnarray}

\noindent (up to addition full time derivative to the Lagrangian).
Generator here is the constraint $\pi_{(ikl)}*\pi_{(ikl)}$,
which, although not written explicitly in Lagrangian, is combination
of constraints of the type $\pi*\pi^{\prime}$ and $\sum\pi$.

Situation for shift of $\phi$ is complicated by occurence of linear
in
$\phi$ terms in the Lagrangian: analogous transformations (\ref{xi})
lead,
as we have seen, to constraints with generalised velocities
$\dot{\pi}$.
On the other hand, since $N^{(3)}_2\geq N^{(3)}_1$, there exist
$N^{(3)}_2\,
-N^{(3)}_1$ relations $f_{\alpha}(|\pi|^2)$ on scalar squares of
$\pi$.
These constraints are consequences of our full set of constraints in
$L_{\rm dual}, L_{\rm scal}$ and are I class constraints generating
transformations (\ref{xi}) for the following particular choice of
parameters:
\begin{equation}
\xi_{(ikl)}=\xi^{\alpha}\frac{\partial f_{\alpha}}{\partial
(|\pi_{(ikl)}|^2)}.
\end{equation}

\noindent Then, up to the full derivative, the following term in the
Lagrangian arises:
\begin{equation}
\Delta_{\xi}L=\frac{1}{2}\dot{\xi}^{\alpha}f_{\alpha}
+\xi^{\alpha}\sum_{(ikl)}\frac{\partial f}{\partial
(|\pi_{(ikl)}|^2)}
\pi_{(ikl)}\circ\frac{D\pi_{(ikl)}}{dt}
\end{equation}

\noindent ($D\pi$ is defined in (\ref{pi-d-pi})). First term is here
combination of constraints. In the second one the differences of
constraints
$f_{\alpha}$ between neighbouring 3-leaves arise. These differences
are
some algebraic constraints on $\pi,~n$ and should be combinations
from our
full set in $L_{\rm dual},~~L_{\rm scal}$ as well.

Thus, the I class constraints are encountered. These are Gaussian one
and
kinematic relations for scalar and dual squares of $\pi$. All other
constraints, apart from those in $L_{\rm dual},~L_{\rm scal}$, should
arise
when varying $L$ in nondynamical variables
$\pi_+,~~n,~~\phi,~~^*\!\phi$.
Since the latter enter $L$ nonlinearly, the equations obtained do not
give,
generally speaking, any constraints on dynamical variables
$\pi_-,~\Omega$,
but rather simply allow one to express nondynamical variables in
terms of
dynamical ones. However, an important exception exists: the scale of
length
of shift-lapse vector at any vertex $i$ enters $L$ linearly.
Therefore,
first, bivectors $n_{ik}$ at this vertex are defined by given
equations
only up to the common scale, and, second, variation in this scale
gives
the above mentioned Hamiltonian constraint (\ref{Ham}) at this
vertex. This
constraint follows by acting on $L$ the following operator:
\begin{equation}
\sum_k n_{ik}\circ\frac{\partial}{\partial n_{ik}}.
\end{equation}

\noindent Substituting into $H_i$ the values of nondynamical
variables in
terms of dynamical ones gives a constraint on $\pi_-,~\Omega$.
Nontrivial
equations of gravity itself arise in Regge calculus at varying edge
lengths,
the Hamiltonian constraint corresponding variation in timelike
edges.
Variation in spacelike and diagonal edges means variation in $\pi_-$
and
gives not the constraints but equations of motion containing time
derivatives.

As for the momentum constraints, these might arise, in analogy with
continuum GR (\cite{Kha2}), by acting on $L$ the operator
\begin{equation}
\pi_{(imk)l}\circ\frac{\partial}{\partial n_{(ik)lm}}-
\pi_{(iml)k}\circ\frac{\partial}{\partial n_{(il)km}}.
\end{equation}

\noindent This operator cancels $L_{\rm dual}$, but now we have also
$L_{\rm scal}$ not cancelled by this operator. As a result, there are
no
analogs of the momentum constraints of continuum GR: shift vectors
enter
$L$ nonlinearly, therefore variation in them allows one only to find
these
themselves.

Thus our system in the space of dynamical variables $\pi_-,~\Omega$
is described by $N^{(3)}_0$ Hamiltonian constraints $H_i$,
$~6N^{(3)}_3$
components of Gaussian constraint $C$ and by additional kinematical
constraints on bivectors $\pi_-$. The I class constraints are
$C,~~\pi*\pi,
{}~~f_{\alpha}(|\pi|^2)$. Since $N^{(3)}_3=2N^{(3)}_2$, it is
convenient to
define each $\pi_-$ in any of two tetrahedrons so that each
tetrahedron
would contain two bivectors defined in it, $\pi$ and $\pi^{\prime}$.
Then
other constraints, a priori II class ones, are $H_i$, $~N^{(3)}_3$
constraints $\pi*\pi^{\prime}$ and $N^{(3)}_3$ functions $g_A$
expressing
scalar products $\pi\circ\pi^{\prime}$ in terms of squares $|\pi|$:
\begin{equation}
\pi\circ\pi^{\prime}=g_{A}(|\pi|^2).
\end{equation}

\noindent It is easy to see that all kinematical constraints
mutually
commute w.r.t. the brackets (\ref{{}}). Nonzero Poisson brackets
arise
only between $H_i$'s in different points and between $H_i$ and
$2N^{(3)}_3$ constraints
$\pi*\pi^{\prime},~~\pi\circ\pi^{\prime}-g_A$.
This means that also $2N^{(3)}_3-N^{(3)}_0$ I class combinations of
functions $\pi*\pi^{\prime},~~\pi\circ\pi^{\prime}-g_A$ exist. On
the
whole, there are $6N^{(3)}_2-N^{(3)}_1-N^{(3)}_0$ I class
constraints.
As $2N^{(3)}_0$ II class ones we can take, in addition to $H_i$,
also
some $N^{(3)}_0$ of products $\pi*\pi^{\prime}$. Without taking into
account
the constraints the number of the degrees of freedom would coincide
with
the number of canonical pairs $6N^{(3)}_2$. Taking into account the
constraints we get this number coinciding with the number of edges
$N^{(3)}_1$. This should be expectable since, generally speaking,
change
of the length of any edge means change of geometry of 3-leaf.

We are faced also with some peculiarity connected with that
Hamiltonian
constraint is II class one. As a result, the length of shift-lapse
vector
$N$ being Lagrange multiplier at this constraint in empty space is
zero.
However, in the presence of matter this singularity dissappear. For
example, contribution of electromagnetic field $F_{\mu\nu}$ into
action
containes the terms of the form
\begin{equation}
g^{00}g^{\alpha\beta}F_{0\alpha}F_{0\beta}V,
\end{equation}

\noindent where $V$ is the volume of 4-simplex, $g_{\mu\nu}$ is
metric.
Since $V\sim N,~~g^{00}\sim N^{-2}$, the given terms are proportional
to
$N^{-1}$, so that equations of motion give strictly nonzero value of
$N$.
One can say that the matter fields prevent the collapse in time axis
by
developing the pressure from within the 4-simplices.

Vanishing the timelike lengths in empty space leads also in some
sence
to triviality of classical dynamics in this space. Indeed, in this
case
Hamiltonian reduces to linear combination of I class constraints.
Since all
these commute with $\pi\circ\pi$, the areas as well as links do not
vary in
time. However, normalised bivectors $n/|n|$ have quite complex
dynamics.
This means that parameters of embedding the 3-leaf into 4-manifold
have
a nontrivial dynamics.

\begin{center}{\bf 5.CONCLUSION.}\end{center}

Having got Regge calculus in canonical form we can write out puth
integral as formal solution to the canonical quantisation problem
for this theory. The functional integral measure is defined by
volume
element in phase space on hypersurface of constraints of the theory
and contains nonlocal factor which is determinant of the Poisson
brackets
of II class constraints. The latter are original constraints of
theory
plus gauge conditions by the number of original I class constraints.
One of interesting feature of Regge theory is that (in the case of
Euclidean signature) integrations over connections (elements of
SO(4),
not of Lee algebra so(4), as in the continuum theory) are finite and
one
does not need to fix the gauge, that is, to divide by the gauge
group
volume. In this case the measure factor will be defined by simply
the
original II class constraints of the theory (Hamiltonian and
kinematical
ones).

Another, unpleasant feature is that this measure is clearly singular
in
the vicinity of flat manifold for which symmetry group is larger and
classification of constraints changes. Therefore in the vicinity of
flat
space the perturbation theory does not exist.

\begin{center}{\bf ACKNOWLEDGEMENTS.}\end{center}

The author is grateful to I.B.Khriplovich, V.N.Popov and
G.A.Vilkovisky
for useful discussions.

\end{document}